# Stochastic Models of Coalition Games for Spectrum Sharing in Large Scale Interference Channels

Ebrahim Karami and Savo Glisic, Senior Member IEEE

Centre for Wireless Communications (CWC), University of Oulu,

P.O. Box 4500, FIN-90014, Oulu, Finland

Abstract - In this paper, we present a framework for analysis of self organized distributed coalition formation process for spectrum sharing in interference channel for large scale ad hoc networks.

In this approach we use concept of coalition clusters within the network where mutual interdependency between different clusters is characterized by the concept of spatial network correlation. Then by using stochastic models of the process we give up some details characteristic for coalition game theory in order to be able to include some additional parameters for network scaling. Applications of this model are: a) Estimation of average time $\tau$ to reach grand coalition and its variance $\sigma_\tau^2$ through closed form equations. These parameters are important in designing the process in dynamic environment. b) Dimensioning the coalition cluster within the network c) Modelling the network spatial correlation characterizing mutual visibility of the interfering links. d) Modelling of the effect of the new link activation/inactivation on the coalition forming process. e) Modelling the effect of link mobility on the coalition forming process.

## I. INTRODUCTION

Cooperation as a new networking paradigm has been used to improve the performance of the physical layer [1, 2] up to the networking layers [3]. Coalitional games are proved to be a very powerful tool for designing fair, robust, practical, and efficient cooperation strategies in communication networks.

Most of the current research in this field is restricted to applying standard coalitional game models and techniques to study very limited aspects of cooperation in networks. This is mainly due to lack of the literature that tackles coalitional games. In fact, most pioneering game theoretical references, such as [4, 5], focus on noncooperative games, touching slightly on coalitional games within a few chapters.

A coalitional game is formally defined by the pair $(N, v)$, where $N$ is the set of players and $v$ is the coalition value. The most common form of a coalitional game is the characteristic form, whereby the value of a coalition $S$ depends solely on the members of that coalition, with no dependence on how the players in $N \setminus S$ are structured. The characteristic form was introduced, along with a category of coalitional games known as games with transferable utility (TU), by Von Neumann and Morgenstern [6].

In a coalitional game with nontransferable utility (NTU), the payoff that each player in a coalition $S$ receives is dependent on the joint actions that the players of coalition $S$ select. The action space depends on the underlying noncooperative game [7]. The value of a coalition $S$ in an NTU game, $v(S)$, is no longer a function over the real line, but a set of payoff vectors, where each element of a vector represents a payoff that player $i$ can obtain within coalition $S$ given a certain strategy selected by $i$ while being a member of $S$. Given this definition, a TU game can be seen as a particular case of the NTU framework [4].

The class of canonical coalitional games, is the most popular category of games in coalitional game theory.

One application of the canonical games for the study of rate allocation in multiple access channels (MAC), within communication networks is presented in [1, 8, 9]. The models tackle the problem of how to fairly allocate the transmission rates between a number of users accessing a wireless Gaussian MAC channel. In this model, the users are bargaining for obtaining a fair allocation of the total transmission rate available. Every user or group of users (coalition) that does not obtain a fair allocation of the rate can threaten to act on its own, which can reduce the rate available for the remaining users. The game is modelled as a coalitional game defined by $(N, v)$ where N is the set of players, i.e., the wireless network users that need to access the channel, and $v$ is the maximum sum-rate that a coalition $S$ can achieve.

[10] presents two main rules for forming or breaking coalitions, referred to as merge and split. The basic idea behind the rules is that, given a set of players $N$, any collection of disjoint coalitions ($S_1,...,S_l$) can agree to merge into a single coalition $G$, if this new coalition $G$ is preferred by the players over the previous state depending on the selected comparison order. Similarly, a coalition $S$ splits into smaller coalitions if the resulting collection ($S_1,...,S_l$) is preferred by the players over $S$.

In canonical and coalition formation games, the utility or the value of a coalition does not depend on how the players are interconnected within the coalition. However, it has been shown that, in certain scenarios, the underlying communication structure between the players in a coalitional game can have a major impact on the utility and other characteristics of the game [11].

In this paper we present a contribution to model coalition games for spectrum sharing in interference channel for applications in large scale wireless ad hoc networks. The rest of this paper is organized as follows,

System model and coalition formation protocol are presented in Section II. The proposed model for coalition formulation problem is presented in Section III. Optimization of the cluster size is presented in Sections IV and corresponding numerical results are presented in Section V. And finally paper is concluded in Section VI.

## II. THE SYSTEM MODEL: PRELIMINARIES

The network consists of a number of links whose transmitters and receivers are randomly located within a certain area $A$. In the case when all links transmit simultaneously, the signal received by a reference receiver $r$ is given as

$$y_r = h_{rr}x_r + \sum_{i \neq r} h_{ri}x_i + n_r, \quad (1)$$

where the second term in right hand side of (1) is interference received from the other transmitters. SINR and its equivalent maximum transmission rate are calculated as,

$$SINR_r = \frac{g_{rr}P_r}{\sum_{i \neq r} g_{ri}P_i + N_0}, \quad (2)$$

$$R_r = \frac{1}{2}\log_2(1 + SINR_r), \quad (3)$$

where $P_i$ is the transmitted power from transmitter of link $i$, $g_{ir}$ the channel gain between the transmitter of link $i$ and receiver of link $r$ and $N_0$ is the receiver background noise power. In this scenario a given transmitter can *spread* its signal over entire available bandwidth or *share* the bandwidth with other users in the network in certain proportion. For the convenience of the presentation let us assume at the beginning that the mutual channel gains are known. This assumption has significant impact on the coalition forming process and for this reason will be reconsidered later.

The above assumption for communication between the users $i$ and $j$ during the process of coalition negotiation some signalling capacity $R_{ij} = R_c$ is required. In the literature the signalling network is referred to as underlying network [12] and the system performance depends significantly on its capacity. In general more sophisticated algorithms, leading faster to optimum operation of the network, would also require more information to be exchanged between the users which consequently require more signalling capacity.

If the interfering level from other users is too high, the reference user $r$ may consider proposing the coalition $S(r, j)$ to a given user $j$ which consists of sharing the spectra in certain proportion.

Again, for simplicity, let us assume that the spectra will be shared evenly (one half for each). The extension to the case of so called fair partitioning of the value of the game where the pay off to the users is proportional to their contribution to the value of the game (Shapley value) [4] is straightforward. Under these conditions we have

$$SINR_{r \in S} = \frac{g_{rr}P_r}{\sum_{i \in S^c} g_{ri}P_i + N_0}, \quad (4)$$

$$R_{r \in S} = \frac{1}{2 \times 2}\log_2(1 + SINR_{r \in S}) - R_S, \quad (5)$$

where $R_S = R_c$ and $S^c$ is the complementary set of users not belonging to coalition $S$. The signal to interference to noise ratio is now improved because the two users in the coalition do not transmit simultaneously, but the user can use only a half of the available spectra (or time slots). The condition for the coalition formation is that the both users are benefiting from such coalition which can be formally expressed as

$$R_{r \in S} \geq R_r. \quad (6)$$

This simple concept will be elaborated later in more details.

### A. Distributed Stochastic Interference Channel Model

In this section, we generalize the previous concept in order to be used in large scale networks. Like in any other generalization, we will have to give up some details in the system modelling, characteristic for coalition game theory, in order to be able to include some additional system parameters relevant to network scaling. We will assume that every user in the network identifies $N$ the strongest interferers (seen by its receiver) and initiate process to establish coalition with these links in order to reduce the mutual interference. The deviation from this assumption to the case where the user is aware only of the aggregate interference would be to initiate the coalition forming process with randomly chosen interferer.

In *fully correlated network (FCN)*, the same set of interfering signals is observed by all receivers. The strength of the interfering signals observed with different receivers is uncorrelated. This models the case where the cluster size $N$ equals the overall number of terminals in the network $M$.

In *partially correlated network (PCN)*, with correlation

$$\rho = \left(1 - \frac{a}{N}\right), \quad (7)$$

the transmitter of specific user $i$, whose receiver identifies $N$ interfering links, is not observed by $a$ receivers from the set of the interfering links. The strength of the interfering signals observed with different receivers is again uncorrelated. In a real network, in every particular time instance, parameter *a* might be different for each link but in average the statistical channel model assumes that parameter *a* is the same for each link. In a fully connected one hop network with *M* terminals and cluster size *N* with *M>N*, probability distribution function $p(a)$ can be expressed as

$$p = \left(1 - \frac{N}{M}\right), \quad (8)$$

$$p(a) = \binom{N}{a} p^a (1-p)^{N-a}. \quad (9)$$

In (8), *p* represents the probability that for the two links *i* and *j*, the receiver of link *i* observes the interference from transmitter of link *j* but the receiver of link *j* does not see the interference from transmitter of link *i*.

## B. Coalition Formation Protocol (CFP)

Once getting the right to initiate coalition formation process, and known channel coefficients, the user *i* proposes the coalition formation to the link *j* with the interfering level
a) higher than the level of its useful signal.
b) maximum interfering level.
c) interfering level with no more than $\delta$ positions below its own received signal level on the ordered list of the received signals.
d) to the arbitrary chosen link (when only the aggregate level of interference is known).
The interfering link *j* will accept the coalition if the interfering level from the user *i* is
e) higher than the level of its useful signal.
f) its maximum interfering level.
g) no more than *δ* positions bellow the level of the received signal level on the ordered list of received signals.
h) The interfering link *j* accepts the coalition conditionally, initiates the channel sharing mode, checks the effects of the coalition forming and confirm or reject the coalition based on the experienced value obtained by the coalition.

## III. CFP MODELLING

The graph presentation of the CFP is shown in Fig.1. Assume there are *M* available links and the process starts with a group of *N* randomly selected users $G(N)$ in singleton status where $N \ll M$. A given link *i* creates an ordered list of interferers $L(i) = \{l(i,j)\} = \{l(i,1), l(i,2), ..., l(i,N)\}$ including its own signal by allocating the lowest index *j=1* to the weakest signal level. Node 1, on graph in Fig.1a, represents the case when the useful signal level is the lowest on the list i.e. $l(i) = l(i,1)$. In the statistical interference channel model, this will happen with probability $p_1 = p(N) = 1/N$. Under the condition represented by node *1* and option a) in Section II.C the link will propose the coalition formation to the link *k* with the interfering level $l(i,k)$ higher than the level of its useful signal that would happen with probability one for *N >1* since all other links have higher level $l(i) = l(i,1) \leq l(i,k)$; for $\forall k$.

Link *k* will have its own list $L(k) = \{l(k,j)\} = \{l(k,1), l(k,2), ..., l(k,N)\}$ and its own level $l(k) = l(k,n)$ can be with same probability $\frac{1}{N}$ anywhere on the list $L(k)$. The level of the link *i*, proposing the coalition, is also uniformly distributed on the list $L(k)$. Extension to the case when these probabilities are arbitrary distribution functions is straightforward. Link *k* will accept the coalition if the level of the link proposing the coalition is higher than its own level on the list $L(k)$. This will occur with probability

$$p_c^{FCN}(N) = p\{l(k,i) > l(k,n) \text{ for } n = 0,1,...,i-1, i+1,...,N\} = \sum_{n=1}^{N} p_n p_{i>n|i \neq n} \quad (10)$$

and consequently,

$$p_c^{FCN}(N) = \begin{cases} \sum_{n=1}^{N} \frac{N-n}{N(N-1)} = \frac{1}{2} & \text{for } N \geq 2 \\ 0 & \text{for } N = 1 \end{cases}. \quad (11)$$

*In partially correlated network (PCN)*, with correlation factor $\rho = (1 - a/N)$, the transmitter of the specific link *i*, whose receiver identifies *N* interfering links, is not seen by *a* receivers from the set of the interfering links. In this case (11) should be modified as

$$p_c^{PCN}(N) = p\{l(k,i) > l(k,n) \,\&\, l(k,i) \in L(k)$$
$$\text{for } n = 0,1,..., i-1, i+1,..., N\} = \sum_{n=1}^{N} p_n p_{i>n|i \neq n} \quad (12)$$

In right hand side of (12), *a* unobservable interferers are excluded from the list $L(k)$ and consequently,

$$p_c^{PCN(a)}(N) = \begin{cases} \sum_{n=1}^{N-a} \frac{N-a-n}{N(N-1)} = \frac{(N-a)(N-a-1)}{2N(N-1)} & \text{for } N \geq a+1 \\ 0 & \text{for } N \leq a \end{cases}. \quad (13)$$

By averaging (13) over *a* using (9), we have

$$p_c^{PCN}(M,N) = \begin{cases} \sum_{a=0}^{N} \binom{N}{a}\left(1-\frac{N}{M}\right)^a \left(\frac{N}{M}\right)^{N-a} \frac{(N-a)(N-a-1)}{2N(N-1)} & \text{for } N \geq 2 \\ 0 & \text{for } N = 1 \end{cases} \quad (14)$$

After some manipulations (14) is simplified as

$$p_c^{PCN}(M,N) = \begin{cases} \frac{1}{2}\left(\frac{N}{M}\right)^2 & \text{for } N \geq 2 \\ 0 & \text{for } N = 1 \end{cases}. \quad (15)$$

When the coalition request is accepted, the process in Fig. 1b will move to the level $G(N-1)$ which represents the same cluster of links where two links have created coalition that will be represented by coalition head which will be negotiating the further coalition process on behalf of the coalition. Effectively from that point on there will be $N$-1 negotiators.

Depending on the capacity of the underlying network, that determines the amount of information that can be distributed about the coalition, the level of interference caused by the coalition will be represented either by the level of the coalition head or the sum of the levels of the coalition members. If the proposal for coalition is not accepted, which occurs with probability $p_c^{'}(N) = 1 - p_c(N)$, the process will remain on level $G(N)$ and the proposals for coalition will be initiated from the same level. In this case, coalition cluster is updated by substituting the link that has rejected the coalition request with a new one from $M$-$N$ remained links.

In the node 1 of the graph in Fig. 1, the level of the user $i$ on list $L(i)$ will not be the lowest with probability $p^{'}(N) = 1 - p_1 = 1 - p(N)$ and the process will move to node 2, which represents the hypothesis that this level is the second on the list given that it is not the lowest. In general, node $k$ represents the probability that the level of the useful signal is the $k$-th on the list given that it is not lower that $k$. Under these assumptions, the transition probabilities are represented in the graph.

Summing up all possible ways to go from $G(N)$ back to $G(N)$ we have the transition probability $p(N, N)$ and to move to $G(N-1)$ the transition probability $p(N, N-1)$ as

$$p(N,N) = \sum_{k=0}^{N-1} p(N-k) p_c^{'}(N-k) \prod_{i=0}^{k-1} p^{'}(N-i) = \frac{1}{N} \sum_{k=0}^{N-1} p_c^{'}(N-k) \quad (16)$$

$$p(N,N-1) = \sum_{k=0}^{N-1} p(N-k) p_c(N-k) \prod_{i=0}^{k-1} p^{'}(N-i) = 1 - p(N,N) \quad (17)$$

and the graph from Fig. 1a can be replaced by an equivalent graph from Fig. 1b, where $z$ represents a fixed time $T$, in z transform, needed for initiation and decision on coalition forming.

By substituting (11) in (16) and (17), transient probabilities for a FCN is simply derived as

$$p^{FCN}(N,N) = \frac{N+1}{2N}, \quad (18)$$

$$p^{FCN}(N,N-1) = \frac{N-1}{2N}. \quad (19)$$

From (18) we can see $p^{FCN}(N, N)$ is bounded as

$(for\ N >> 1) \quad 1/2 < p^{FCN}(N,N) \le 1 \quad (for\ N = 1). \quad (20)$

And for a PCN which is a more general model, transient probabilities are computed as follows,

$$p_M^{PCN}(N,N) = \begin{cases} 1 - \frac{1}{2N} \sum_{k=0}^{N-2} \left(\frac{N-k}{M-k}\right)^2 & for\ N \ge 2 \\ 1 & for\ N = 1 \end{cases} \quad (21)$$

$$p_M^{PCN}(N,N-1) = \begin{cases} \frac{1}{2N} \sum_{k=0}^{N-2} \left(\frac{N-k}{M-k}\right)^2 & for\ N \ge 2 \\ 0 & for\ N = 1 \end{cases} \quad (22)$$

From (21), upper and lower bounds of $p^{PCN}(N, N)$ are as

$$1 - \frac{N^2}{4M^2} \le p_M^{PCN}(N,N) \le 1 - \frac{(N-1)N}{2M^2} \quad (23)$$

In (23) upper bound holds for $N$=1 and lower bound holds for $N$=2.

Both equations (18) and (21) for $N$=1, give the same result i.e. $p^{FCN}(1,1) = p^{PCN}(1,1) = 1$ which is an obvious result because in this case there is no remaining coalition

IV. OPTIMIZATION OF THE CLUSTER SIZE

The size of the cluster $N$ is the parameter to be optimized. This optimization is performed through maximization of the throughput received by the cluster head. Using (3) and (4) and definition of the coalition formation process, for the network with $M$ links, available rate for a singleton with index 0 is

$$R_0 = \frac{1}{2} \log_2\left(1 + \frac{S}{N_0 + \sum_{i \ne 0} I_i}\right). \quad (24)$$

If we assume that the same rate $R_c$, is dedicated for communication between the cluster head with each coalition candidate and that the spectra share is the same for any member of the coalition, the rate per member is

$$R_N = \frac{1}{2N} \log_2\left(1 + \frac{y}{1 + \sum_{i \in S^c} y_i}\right) - (N-1)R_c, \quad (25)$$

where $y$ is SNR of useful signal (signal of coalition head) and $y_i$ is SNR of $i$th interfering signal. At the beginning, $S^c$ is unknown and therefore the interference term $\sum_{i \notin S^c} y_i$ is estimated as follows,

$$\sum_{i \notin S^c} y_i = \sum_{i=1}^{M} y_i - \sum_{i \in S^c} y_i = M\bar{y} - \sum_{i \in S^c} y_i, \quad (26)$$

Since interference coming from the member of coalition cluster is greater than useful signal term $\sum_{i \in S^c} y_i$ can be estimated as,

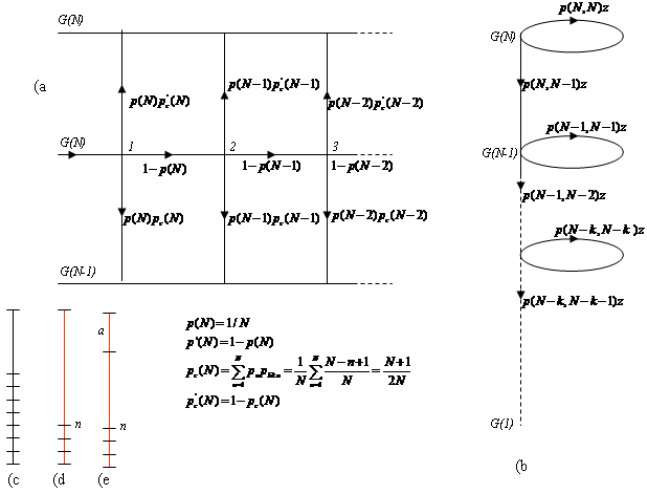

Fig.1. Graph presentation of CFP.

$$\sum_{i \in S^c} y_i \cong y + (N-1)E(y_i | y_i > y) = y + (N-1)\frac{\int_y^\infty xP(x)dx}{\int_y^\infty P(x)dx}. \quad (27)$$

where $P(.)$ is probability distribution function for $y_i$ s which is dependent on the propagation environment and distribution of nodes. For getting rid of $P(y)$, we approximate right hand side of (27) as

$$E(y_i | y_i > y) = y + \frac{\int_o^\infty xP(x+y)dx}{\int_0^\infty P(x+y)dx} \cong y + \bar{y}. \quad (28)$$

Consequently

$$\sum_{i \in S^c} y_i \cong Ny + (N-1)\bar{y}. \quad (29)$$

From (25), (26), and (29) the rate per member as a function of $N$ is expressed as

$$R_N = \frac{1}{2N}\log_2\left(1 + \frac{y}{1+(M-N+1)\bar{y}-Ny}\right) - (N-1)R_c. \quad (30)$$

Equation (30) shows that by increasing the $N$, not only the number of interferers decreases, but also the average of received interference from each interferer decreases and this is the direct result of selection of the candidates with stronger interference.

Extension to the fair distribution of the game value, proportional to the contribution to the gain is straightforward. The optimum $N$ is calculated through the maximization of sum throughput. For illustration purposes as an example assume that $R_c$ is proportional to the available data rate as

$$R_c = \alpha R = \frac{\alpha}{2}\log_2\left(1 + \frac{y}{1+(M-N+1)\bar{y}-Ny}\right). \quad (31)$$

Then $R_N$ is derived as

$$R_N = \frac{1-\alpha N(N-1)}{2N}\log_2\left(1 + \frac{y}{1+(M-N+1)\bar{y}-Ny}\right), \quad (32)$$

where $\bar{y}$ is the average signal to noise ratio in $S^c$. In this case optimum $N$ is the closest integer to the solution of the following equation $\partial R_N / \partial N = 0$ resulting into

$$Ln\left(1 + \frac{y}{1+M\bar{y}-N(y+\bar{y})}\right) = \frac{N(1-\alpha N)y(y+\bar{y})}{(1+\alpha N^2)(1+(M-N+1)\bar{y}-Ny)(1+(M-N+1)\bar{y}-(N-1)y)}. \quad (33)$$

In the case where $R_c$ is constant optimum, $N$ is the closest integer to the solution to the following equation

$$R_c Ln(2) = \frac{y(y+\bar{y})}{(1+(M-N+1)\bar{y}-Ny)(1+(M-N+1)\bar{y}-(N-1)y)} \\ - \frac{1}{2N^2} Ln\left(1 + \frac{y}{1+(M-N+1)\bar{y}-Ny}\right). \quad (34)$$

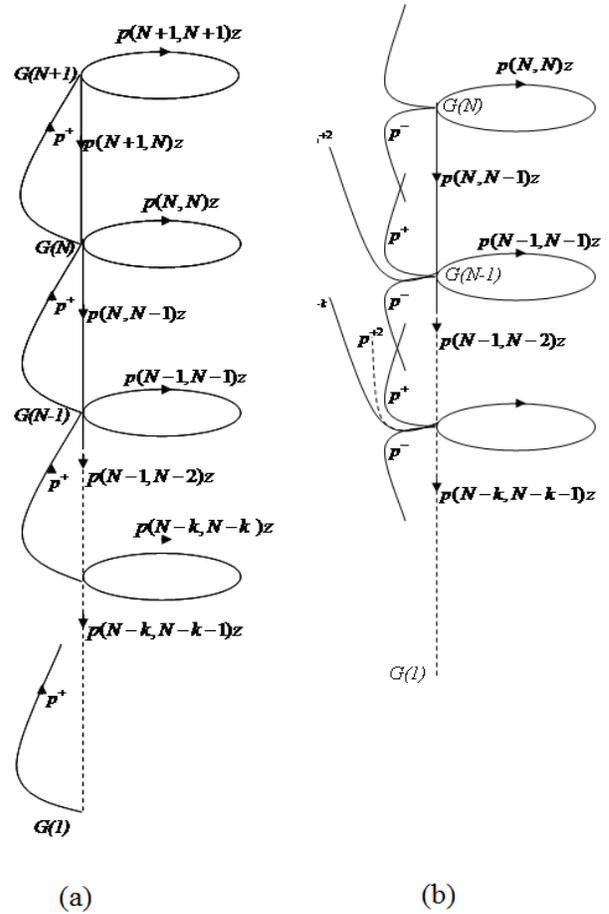

Fig. 2. Modelling of the a) arrivals of a new link b) departure of a link during the coalition formation process.

## V. NUMERICAL RESULTS

In this Section, we present numerical results for the proposed model. All nodes are assumed uniformly distributed on a rectangular and results are averaged over 20000 independent runs. Figs. 3-4 presents throughput and average optimum cluster size w.r.t. SNR for both fix and proportional $R_c$ and compare throughput achieved by them with no coalition case i.e. when $N$=1. It is clear that comparing the results of fix and proportional $R_c$ does not make sense but this Fig. is presented to compare them with no coalition case.

Fig. 3 presents optimum cluster size for $M$=10 and 25. For low SNR values when the performance is rather dependent to SNR than interference, coalition does not much improve the performance and therefore optimum cluster size is small. On the other hand, for large values of $R_c$ and $\alpha$, optimum cluster size is small because larger cluster sizes need more dedicated bandwidth for coalition negotiation. Fig. 4 confirms this result and we can see for large values of $R_c$ and $\alpha$ coalition does not much improve the throughput. In a practical network, when nodes are stationary the coalition negotiation is performed only once and when we have a dynamic network, further negotiations are necessary upon mobility in the networks and actually the average required $R_c$ is low. From Fig. 3, we can see, for very small values of $R_c$ and $\alpha$ the average optimum cluster size is almost $M$/2.

## VI. CONCLUSION

In this paper, we model the self organized distributed coalition games for spectrum sharing. The proposed model is efficient for large scale *ad hoc* network especially in dynamic environments when the number of active links is varying. Then using the proposed model, mean and variance of the required time for the coalition process are analytically derived in closed form. Throughput as a function of cluster size is analytically derived and the cluster size is optimized to achieve the maximize throughput. Simulation results prove the throughput gain compared to no coalition case.

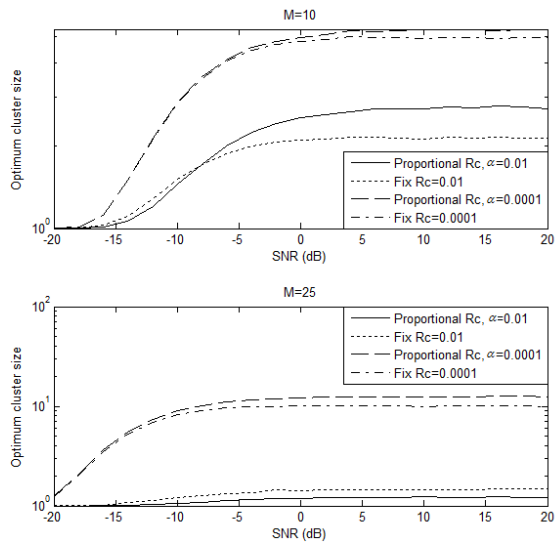

Fig. 3. Optimum cluster size w.r.t. SNR.

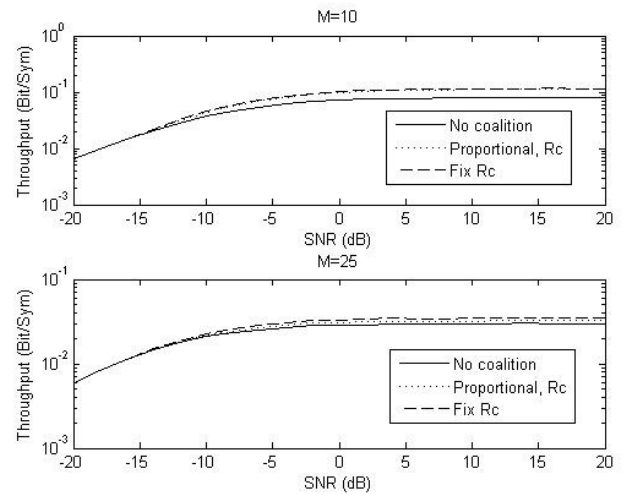

Fig. 4. Throughput w.r.t. SNR when cluster size is optimized for $\alpha = R_c = 0.01$.